%% file: main.tex
\documentclass[runningheads]{llncs}
\usepackage{lipsum}
\usepackage{balance}
\usepackage{enumitem}
\usepackage[dvipsnames,svgnames]{xcolor}
\usepackage{cite}
\usepackage{verbatim}
\usepackage{arydshln}
\usepackage{booktabs}
\usepackage{threeparttable}
\usepackage{multirow}
\usepackage[utf8]{inputenc}
\usepackage{url}
\usepackage{amssymb}
\usepackage{bbding}
\usepackage{array} % required for text wrapping in tables
\usepackage{pifont}
\usepackage{wasysym}
\usepackage{utfsym}
\usepackage{fontawesome}
\usepackage{amsfonts}
\usepackage{algorithmic}
\usepackage{graphicx}
\usepackage{stfloats}
\usepackage{textcomp}
\usepackage{lscape}
\usepackage{color}
\usepackage{hyperref}
\usepackage{subcaption}
\def\BibTeX{{\rm B\kern-.05em{\sc i\kern-.025em b}\kern-.08em
    T\kern-.1667em\lower.7ex\hbox{E}\kern-.125emX}}
\usepackage{graphicx}
\definecolor{dgreen}{RGB}{0,150,0}
\definecolor{dred}{RGB}{200,0,0}

\newcommand{\ie}{\emph{i.e., }}

\newcommand{\etc}{\emph{etc.}}

\newcommand{\maRes}[2]{\textcolor{dgreen}{#1} (\textcolor{dred}{+#2})}
\usepackage[T1]{fontenc}

\begin{document}
%
%\title{CTEKG: Constructing Complex Threat Event Knowledge Graph from Cyber Threat Intelligence}
%\title{AttacKG\textbf{+}: Extracting Attack Knowledge Graphs with Large Language Models}
\title{AttacKG\textbf{+}: Boosting Attack Knowledge Graph Construction with Large Language Models}

%
%\titlerunning{Abbreviated paper title}
% If the paper title is too long for the running head, you can set
% an abbreviated paper title here
%
\author{Yongheng Zhang \inst{1} \and
Tingwen Du\inst{2} \and
Yunshan Ma\inst{3} \and Xiang Wang \inst{2} \and Yi Xie \inst{1} \and Guozheng Yang \inst{1} *   \and Yuliang Lu \inst{1} \and Ee-Chien Chang \inst{3}\\
\email{yangguozheng17@nudt.edu.cn}}
\authorrunning{F. Author et al.}
% First names are abbreviated in the running head.
% If there are more than two authors, 'et al.' is used.
%
\institute{College of Electronic Engineering, National University of Defense Technology\and
University of Science and Technology of China\and
National University of Singapore\\
}

\maketitle              % typeset the header of the contribution
\begin{abstract}
%% Text of abstract
%Threat event extraction aims to extract security events in historical cyber threat intelligence, transforming unstructured information into a structured form, and portraying cyber security events' evolutionary trend.
     Attack knowledge graph construction seeks to convert textual cyber threat intelligence (CTI) reports into structured representations, portraying the evolutionary traces of cyber attacks.
     Even though previous research has proposed various methods to construct attack knowledge graphs, they generally suffer from limited generalization capability to diverse knowledge types as well as requirement of expertise in model design and tuning.
     Addressing these limitations, we seek to utilize Large Language Models (LLMs), which have achieved enormous success in a broad range of tasks given exceptional capabilities in both language understanding and zero-shot task fulfillment.
     Thus, we propose a fully automatic LLM-based framework to construct attack knowledge graphs named: AttacKG\textbf{+}. Our framework consists of four consecutive modules: rewriter, parser, identifier, and summarizer, each of which is implemented by instruction prompting and in-context learning empowered by LLMs.
     Furthermore, we upgrade the existing attack knowledge schema and propose a comprehensive version. We represent a cyber attack as a temporally unfolding event, each temporal step of which encapsulates three layers of representation, including behavior graph, MITRE TTP labels, and state summary. 
     Extensive evaluation demonstrates that: 1) our formulation seamlessly satisfies the information needs in threat event analysis, 2) our construction framework is effective in faithfully and accurately extracting the information defined by AttacKG\textbf{+}. and 3) our attack graph directly benefits downstream security practices such as attack reconstruction.
     All the code and datasets will be released upon acceptance.

\end{abstract}

\input{1_Introduction}
\input{2_Task_Defination}
\input{3_Approach}

\input{4_Evaluation}
\input{5_Related_Work}

\input{6_Conclution_and_Future_work}

\begin{credits}
% \subsubsection{\ackname}The project was supported by Open Fund of Anhui Province Key Laboratory of Cyberspace Security Situation Awareness and Evaluation.

\subsubsection{\discintname}
The authors have no competing interests to declare that are relevant to the content of this article. 
\end{credits}

\input{ref}
\input{7_Appendix}

\end{document}

%% file: 1_Introduction.tex
\section{Introduction}
\label{introduction}

% Attack knowledge graph task introduction
To better combat the intensifying threat of advanced cyber attacks, security analysts are exchanging cyber threat intelligence (CTI) to enhance their awareness and respond effectively to threat events.
Since CTI reports are written in the free form format of natural language, the researchers propose the task of attack knowledge graph construction, aiming to structurally analyze the cyber attack process~\cite{AttacKG}.
Due to the significant values in both academia and industry, attack knowledge graph construction has garnered growing attention in recent years~\cite{EXTRACTOR,ThreatKG,TTPDrill}.
% Due to the important research and application value of attack knowledge graph construction, it has received more and more attention in the field of network security analysis.
% identifying attack behaviors requires analyzing the semantics of unstructured CTI texts. 

% The specialized nature of cybersecurity knowledge makes this task quite challenging.

% Researchers have introduced knowledge-enhanced attack graphs for structured summarization of CTI reports, which reduces the difficulty of understanding threat semantics and accelerates the content parsing of intelligence.

% Limitations of current work
Several attack knowledge graph construction methods have been proposed by researchers.
EXTRACTOR~\cite{EXTRACTOR} and THREATKG~\cite{ThreatKG}  extract threat entities and relations by constructing domain regular expressions or ontology models, thus transforming unstructured text written in natural language into structured knowledge graphs. 
Further, in order to normalize the technical connotations of the description of the attack process,
TTPDrill~\cite{TTPDrill} and AttacKG~\cite{AttacKG} propose to leverage the MITRE TTP (Tactics, Techniques, and Procedures) ontology~\footnote[1]{\url{https://attack.mitre.org/}} to tag the structured KG.
Nevertheless, current methods for attacking knowledge graph construction have two main limitations:

% limitation-1
\textbf{Limitation 1}: The current construction model has limited semantic understanding capability, leading to poor generalization to diverse and newly emerging attack scenarios and knowledge types. The capability of existing threat information extraction methods is extensively restricted by limited volume of training data as well as relatively small model size. As a result, the model is difficult to generalize to various open scenarios and cannot cover a wide range of security knowledge types~\cite{Nguyen,zhao-etal-2018-document,Lishuang,Kodelja}, resulting in major threat information loss when faced with unknown security knowledge that cannot be understood and identified.

% limitation-2
\textbf{Limitation 2}: Existing methods substantially depend on the specialized design and meticulous curation of natural language processing or graph matching models, thereby posing a challenge for a majority of security practitioners lacking proficiency in these domains. To achieve better information extraction results for the target object, existing methods require a lot of manpower to fine-tune the model. This process requires technicians with deep experience in debugging model parameters.
% and a relevant understanding of information characteristics in the cybersecurity domain. 
This problem becomes a hindrance for those cybersecurity technicians with weaker background knowledge in the AI domain to deeply participate in the design of cyber threat intelligence extraction methods.

% Task Adaptability of Large Language Models点击并应用
The breakthrough of Large Language Models (LLMs)~\cite{LLM1,LLM2} has shed light on these problems. 
First, LLMs use massive open knowledge data in the pre-training process, so it has a powerful contextual understanding and knowledge reasoning ability to understand various domains and various kinds of knowledge.
Second, LLMs can perform a wide variety of zero-shot and few-shot tasks by means of instruction following and in-context learning, without requiring special model structure design or training on specific datasets.
Therefore, extracting attack knowledge graphs with LLMs can be a good solution to the two limitations mentioned above.
Currently, LLMs have been explored for initial applications in the field of cyber security~\cite{wrsch2023llms,ferrag2023revolutionizing,microsoft,qianxin}. But in terms of cyber KG construction, it's still under exploration.

% Introducing the framework
Leveraging LLMs for the attack knowledge graph construction, we propose a fully automatic framework with four modules: rewriter, parser, identifier, and summarize, each of which is implemented with dedicated prompt engineering and in-context learning based on LLMs. The rewriter filters out redundant information and organizes report content into sections, each corresponding to a tactical stage defined in the MITRE TTP and maximally preserving the key knowledge. Thereafter, given the rewritten sections, the parser extracts the behavior graph, including the atomic event triplets, temporal relations between threat actions, as well as entity-entity relations. Next, given the behavior graph and the rewritten sections, the identifier then matches the behavior graph and the rewritten sections to the appropriate MITRE technique labels. Finally, the summarizer summarizes the situation and state at the end of each tactical stage. 
% 知识scheme
Moreover, given the superior capability of LLMs in understanding and extracting various cyber knowledge, we propose a more comprehensive schema by synthesizing existing works, encapsulating multiple levels of threat knowledge: threat behavior, TTP labels, and state summary. Hence, our schema is versatile to cover richer threat information. 
We name our approach as AttacKG\textbf{+}, on the one hand, to highlight our contributions in boosting the performance of attack knowledge graph construction, on the other hand, to give credit to the pioneering work AttacKG~\cite{AttacKG} and expect further contributions to this valuable problem. 
% We named this new knowledge schema: AttacKG\textbf{+}.

 We implement AttacKG\textbf{+} and evaluate it against 234 techniques of 14 tactics crawled from MITRE and 500 CTI reports collected from multiple intelligence sources~\cite{Cisco,Microsoft}. AttacKG\textbf{+} successfully identifies 7,305 technique instances, 20,350 entity instances, and 10,175 relation instances.
 Our experimental results show that AttacKG\textbf{+} significantly outperforms existing CTI parsing solutions such as EXTRACTOR~\cite{EXTRACTOR} and AttacKG~\cite{AttacKG} in open scenarios: the F-1 score is significantly improved in the threat entity/relation extraction task and the technique identification task.
 The qualitative evaluation of the tactical rewrite report is very effective, and the drawn AttacKG\textbf{+} shows the attack process very well. Moreover, our method is easy to use, without requiring any prior knowledge of natural language processing and graph learning, and friendly to a broad range of security practitioners. 
 The main contributions are as follows:

\begin{itemize}[leftmargin=*]
      \item To the best of our knowledge,  we are among the first to explore LLMs to facilitate the task of attack knowledge graph construction, forming a novel, general, and user-friendly paradigm. 

      \item We propose a fully automated LLM-based framework AttacKG+, while introducing an upgraded multi-layer knowledge schema for cyber threats.

      \item Extensive evaluation demonstrates the advantage of AttacKG\textbf{+} and the effectiveness of the construction framework. As a by-product, we construct two datasets from 500 CTI reports, \ie Re-CTI and CTI-TE.

\end{itemize}

%% file: 2_Task_Defination.tex
\section{Task Defination}
\label{Task Defination}

Current research still cannot fully reflect the threat event landscape: First, current efforts lack the tactical slicing of CTI reports and therefore cannot delineate the stages of threat events. This results in an inability to monitor situational information at all stages of attacks. Second, current threat knowledge extraction work in the various elements are discrete existence, manifested in the extraction of threat entities, technology labels, and other information is not closely coupled, the lack of threat events on the overall extraction program~\cite{Acing,EXTRACTOR,TTPDrill}. Catering to the powerful capabilities of LLMs, we systematically upgrade the existing cyber security KG schema and propose a more comprehensive attack knowledge graph scheme for CTI reports.

\begin{figure*}[h]
    \centering
    \includegraphics[width=0.92\linewidth]{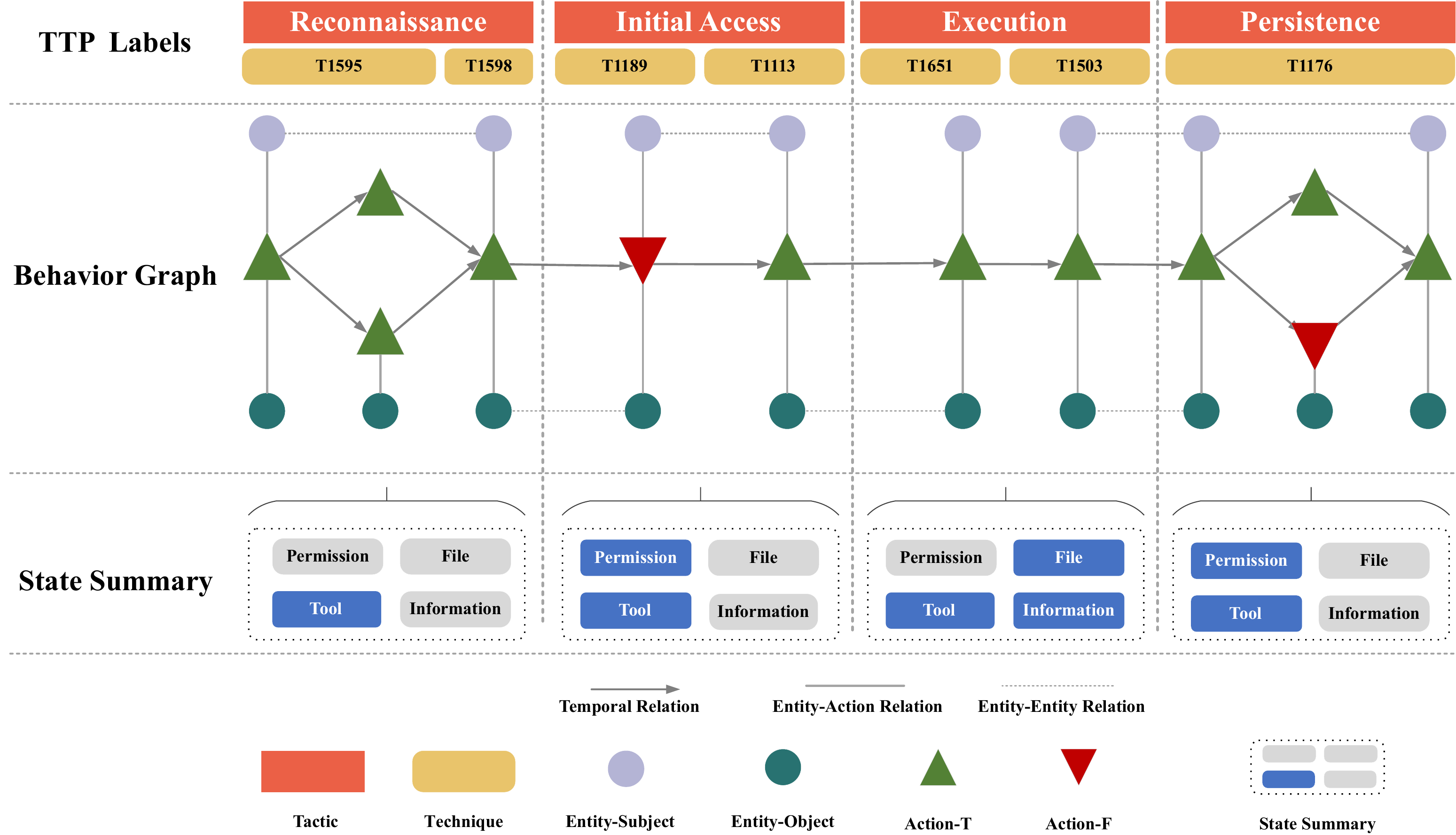}
    \caption{The overview of AttacKG\textbf{+}'s knowledge scheme. We formulate it as a temporally unfolding complex event, each temporal step of which consists of three layers of representation: \textbf{Beharivor Graph} [mid-layer], \textbf{TTP Labels} [top-layer], and \textbf{State Summary} [bottom-layer].
   }
    \label{AttacKG}
\end{figure*}

As shown in Figure~\ref{AttacKG}, we formulate a cyber attack as a temporally unfolding complex event, each temporal step of which is described by a three-layered representation, encapsulating behavior graph, TTP labels, and state summary. 
Specifically, as the backbone of the threat event, behavior graph details the attack process with a structured representation. It consists of a set of temporally connected atomic events, each represented as a triplet $(s, a, o)$, where $s$, $a$, and $o$ stand for the subject, action, and object, respectively. The temporal relation between actions is implied by a directed edge, while there are also non-action entity-entity relations.
On top of behavior graph, we identify the involved tactic and technique patterns based on the TTP matrix provided by MITRE and align TTP labels with its corresponding behavior sub-graphs. This process helps us to better understand the technical connotations of threat events.
In addition, we incorporate a state summary layer to complement threat information by monitoring the changing state of permissions, tools, files, and information within different tactical stages. This information leads to a more complete characterization of the changing cybersecurity situation in each stage.

\vspace{-0.6cm} 
\vspace{-0.8cm}

%% file: 3_Approach.tex
\section{Approach}    
     %In this section, we introduce a threat event extraction framework for cyber threat intelligence based on prompt engineering, describing its general framework as well as the design ideas of each module.
     In this section, we introduce a fully automatic LLM-based framework for AttacKG\textbf{+} construction from CTI reports. We first present the overall design of this framework and then describe the details of each key module.

\begin{figure*}[htbp]
    \centering
    \includegraphics[width=1\linewidth]{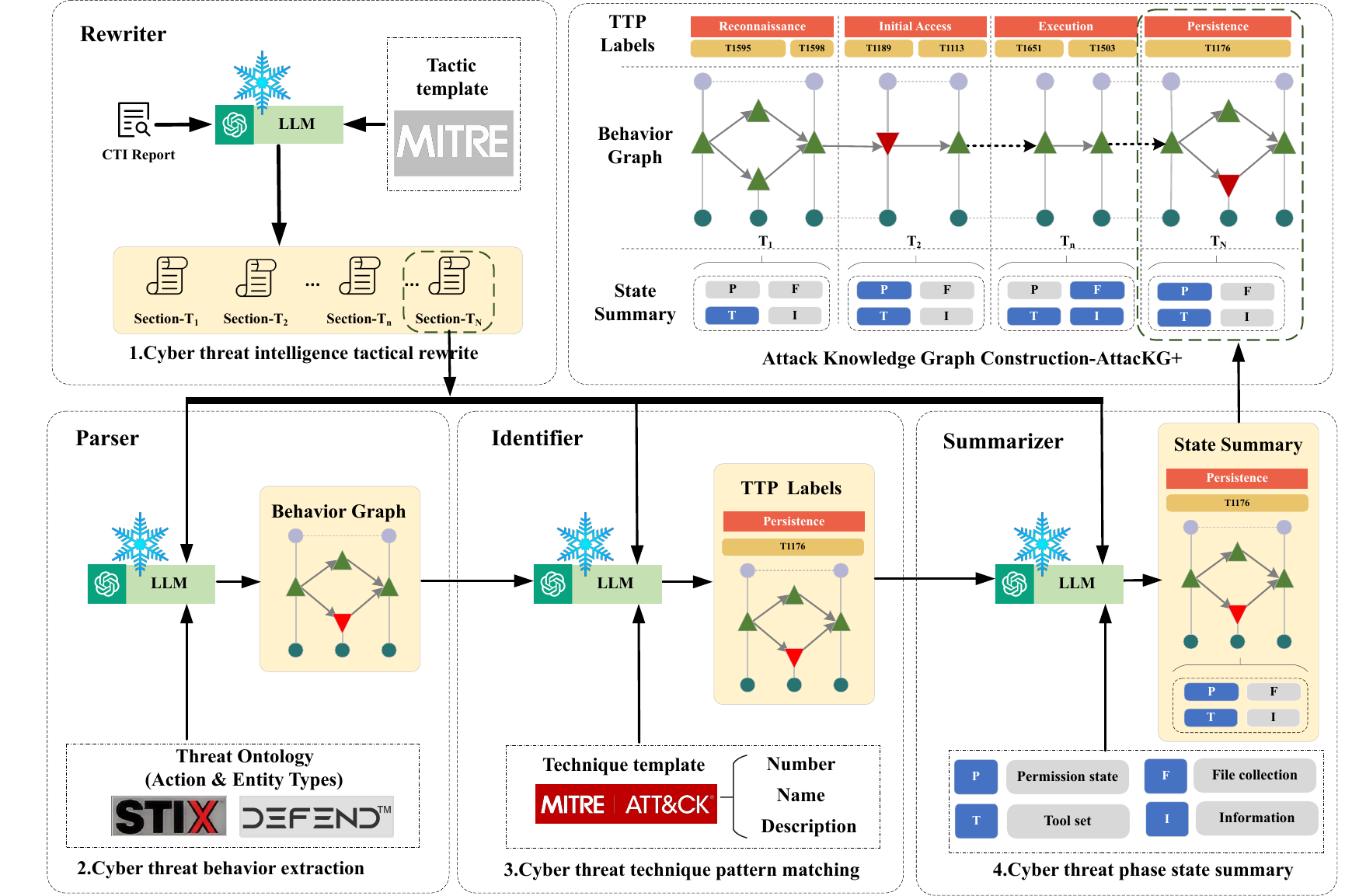}
    \caption{The overall framework of our AttacKG\textbf{+} consists of four components: 1) \textbf{Rewriter}, which enables tactical rewriting of threat intelligence; 2) \textbf{Parser}, which extracts threat entities and relations from threat intelligence; 3) \textbf{Identifier}, which identifies patterns of attack techniques used in cyber threat intelligence; and 4) \textbf{Summarizer}, which performs a stage-by-stage situational summary of threat intelligence.
}
    \label{overview}
\end{figure*}

%\subsection{Overview of CTEKG}
     As shown in Figure~\ref{overview}, the attack knowledge graph construction framework for cyber threat intelligence is composed of four modules: \textbf{Rewriter}, \textbf{Parser}, \textbf{Identifier}, and \textbf{Summarizer}. 

\vspace{-0.5cm} 
\vspace{-0.6cm}

\subsection{AttacKG\textbf{+} Threat Extraction}
     In this section, we introduce the information extraction process of the CTI report, which consists of two modules: \textbf{Rewriter} and \textbf{Parser}.
     The objective is to extract threat behaviors from CTI reports and form a threat behavior graph based on their temporal relations.
     
     After analyzing CTI reports, we can find that well-written reports can effectively portray the process of how various types of attacking entities in cyber threat behavior interact with each other to achieve confrontation goals, as well as provide detailed descriptions of the tactics and techniques involved.
     However, since different security vendors and authors, different CTI reports have different features in terms of language style, writing logic, \etc 
 making it still challenging to extract threat behaviors from them accurately.
     Here, we summarize three key challenges:

     \textbf{[C1]Threat Intelligence Noise Processing}.
     Threat reports contain large amounts of irrelevant text. 
     Often, only a small portion of the report describes the attack behavior. 
     For example, a geographic source describing the malware, while interesting, does not contribute to the analysis of the malware's application.
     
     \textbf{[C2]Threat Entity and Relation Extraction}. 
     CTI reports provide an overview of the attack workflow. CTI reports often have complex syntax and semantics, and proper punctuation is sometimes lacking in these reports. 
     These can easily affect the extraction of threat entities and relations in reports.
     
     \textbf{[C3]Co-reference Resolution}. 
     There are two types of co-references in the CTI report. Explicit co-references use pronouns such as "it" and "they" or the definite article "the", while implicit coreferences use synonyms to refer to entities that appear in the preceding text.

     With prompt engineering, we design a CTI report extraction pipeline to overcome the above challenges (containing both Rewriter and Parser parts in Overview). 
     Rewriter is oriented towards solving [C1], and Parser is oriented towards solving [C2][C3].
     It is worth mentioning that CTI reports published in the field are currently shared as PDFs, so we use relevant open-source tools (such as \textit{pdfpulmer} ~\footnote[2]{\url{https://pypi.org/project/pdfplumber/}}) to convert them to a uniform text format.

\noindent\textbf{CTI Reports Rewrite based on Tactic}.
     Raw CTI reports contain many other classes of information that are weakly correlated with the evolution of security events.
     We focus on the threat behaviors, technique patterns, and stage situations involved in the attack process, and the redundant, weakly relevant information becomes noise in the information extraction process and affects the extraction effect.
     So, we reconstruct the original threat intelligence report based on the TTP matrix's 14 categories of tactical categorization.

     We rewrite the CTI report for two purposes.
     First, data pre-cleaning. Screening out the main content describing the attack process, filtering other redundant information unrelated to the attack process, and improving the information quality of threat intelligence.
     Second, the temporal sequence is pre-composed. The evolution of events in the report is sorted according to the chronological relationship between the tactical classification and the original content, and the sorted content conforms to the standard tactical classification framework.

     Report rewriting aims to reconstruct cyber threat intelligence using tactical labels based on the comprehension and summarization capabilities of LLMs.
     The prompt template can be found in Appendix~A (Rewriting Prompt). 

\begin{itemize}
      \item \textbf{Input Tactical Definitions}. 
      First, we input the definitional information of the tactical labels (name, definition, examples) from MITRE to the LLMs and set the Other label to store the contextual information of non-tactical features.
\end{itemize}

\begin{itemize}
      \item \textbf{Setting Rewrite Rules}. 
      Second, write the prompt template for the rewrite task and set the rewrite rules and output format.
\end{itemize}

\begin{itemize}
      \item \textbf{Perform Rewriting Task}. 
      Finally, we input the text of the original CTI report, perform the rewriting task, and wait for the LLMs to output the rewriting result.
\end{itemize}

     CTI report texts are from different cybersecurity vendors and authors written with very different logic and text organization. Hence they contain a lot of redundant information that has little to do with the evolution of cybersecurity incidents.
     In contrast, the rewritten CTI report has the following characteristics.
     First, the logic of the report is standardized. 
     The text is divided into phases based on the type of tactic to which the content belongs, is hierarchical, and does not contain redundant information.
     Second, the report time sequence is organized. 
     The rewritten text message considers the relative temporal relation of threat behaviors in the content reconstruction and can portray the attacker's invasion thought process according to the sequential flow of tactical implementation.

\noindent\textbf{CTI Reports Parser}.
     To achieve malicious purposes, attackers often take a series of threatening behaviors in conjunction with the invasion environment, penetrate layer by layer, and gradually approach the final target.
     Therefore, extracting threat actions is fundamental in studying the evolution of cybersecurity events. Concerning complex event evolution, the formula for threat atomic events as a triplet $( s, a, o)$ in Section \ref{Task Defination}.
     % , where $s$, $a$, and $o$ stand for subject, threat action, and object, respectively.
     % The relevant elements in the threat atom event are shown in Figure~\ref{ThreatAction}.

     Threat actions ($a$) are a type of verb-like relational set used to characterize the sequential relationship between actions in a CTI report.  The green forward triangle symbol denotes a successful execution, while the red reverse triangle symbol indicates a failed execution.  These actions are linked by directed edges, which convey the temporal relation between them.  Typical actions include collecting, sending, and switching.  The subject ($s$) of the action is the initiator, while the object  ($o$) is the target.  Entities represent specific objects in cyber threat intelligence, such as attackers, exploits, and tools, are represented in the cyber threat process.  The entity-entity relations are distributed among non-verbal relationships, such as located-at and belong-to.  A threat action tuple is formed by associating the action subject with the target object and its relations and attributes.
     The chain extraction of threat actions is thought of as follows, with the prompt template referenced in Appendix~A~(Triplet Extractor Prompt).

\begin{itemize}
      \item \textbf{Modeling the Threat Ontology}. 
      Design the ontology model in conjunction with the STIX specification~\cite{STIX} and the DAO specification~\cite{DAO} to set up entity candidate collections and relation candidate collections.
      \item \textbf{Set the Extraction Rules}. 
      Using the threat action as the core of the atomic event, the subjects involved in the action and the implementation objects are extracted separately. 
      Add referential disambiguation qualification to rules to automatically capture pronoun features. 
      At the same time, associating non-verbal relations builds the basic tuple of threatening actions.
      \item \textbf{Determine the Temporal Relation}. 
      Finally, a chained representation of the entire threat action is strung together based on the relative temporal relation of the threat action as it appears in the report rewrite text.
\end{itemize}

\subsection{CTI Reports Identifier for Technique}
     Threat actions are instantiated representations of attackers' efforts to achieve malicious ends but do not enable cybersecurity personnel to understand the technical connotations that follow the actions.
     To assist security analysts in taking targeted mitigation measures at crucial technical nodes to avoid further threat behavior.
     We need to match technique patterns of threat actions to understand better the technical thinking and implementation logic of intrusion behavior.
     
     The template for the technical model used in this paper is derived from the Tactic and Technique modules of the TTP matrix.
     MITRE summarizes the current network technologies and constructs the TTP matrix to standardize the numbering and naming of the technologies and to give program examples.
     Techniques are subcategories under tactics. 
     Under a tactical category, several types of technological means serve the purpose of that tactic.
     For example, under the tactic Defensive Evasion there exists the technique \textit{T1612 -Build Image on Host}.
     The tactical connotation of Defensive Evasion is that "Defensive Evasion consists of techniques adversaries use to avoid detection throughout their compromise". 
     The technical connotation of \textit{T1612} is that  "Adversaries may build a container image directly on a host to bypass defenses that monitor for the retrieval of malicious images from a public registry".

     Our technique matching method is to annotate the tactical rewritten text with technical types by writing prompt templates and utilizing the analysis and matching capabilities of the LLMs, see Appendix~A~(Technique Identifier Prompt) for the prompt templates and the idea of technical matching as follows:

\begin{itemize}
      \item \textbf{Understanding Tacticalized Texts}. 
      Since the rewritten text has been divided according to tactical phases, separate technique matching can be performed for different tactical phases, thus reducing the total length of the tokens of prompt over multiple reports.
      \item \textbf{Importing TTP Technique Templates}. 
      We input the technique numbers, names, and examples from the corresponding tactics into the LLMs to provide them with domain background knowledge.
      \item \textbf{Preparation of Identification Rules}. 
      The output is formatted using the LLMs, with the text serial numbers and their corresponding technique templates labeled according to the technique model.
      \item \textbf{Technique Template Alignment}. 
      Since threat actions and technique templates are extracted or matched based on rewritten text, the alignment mapping of threat action chains to technique templates can be accomplished based on the correspondence of text positions.
\end{itemize}

\subsection{CTI Reports State Summarizer}

     The combined use of several technologies is often for a tactical purpose in cyberspace (such as Reconnaissance, and Defense Evasion).
     Extracting the state summary information of the intrusion scenarios according to the tactical phases can effectively determine the tactical value of the current phase and the security state of the environment. 
     The process of state summary can be divided into three parts, with the help of the prompt template provided in Appendix~A~(State Summarizer Prompt).
     
\begin{itemize}

      \item \textbf{Defining the Framework of Summary}. 
      Sorting through the contents of the summary, we select four areas of information:
      1) permission state, describing the current stage of the attacker's acquisition of privileges from the victim;
      2) file collection, describing the files that the attacker has obtained from the victim at this stage;
      3) information, describing the current stage in which the attacker has obtained sensitive information from the victim;
      and 4) tool set, scripts or tools used by the attacker in the current phase.
      
      \item \textbf{Dividing the Summary Stage}. We use the tactical phases of the tacticalized rewrite of the text as the basis for dividing each phase, and for each phase of tactical action extracted the relevant elements of the summary paper framework.
      
      \item \textbf{Implementation Summary Extraction}. Finally, we constructed the above extracts and delineation criteria as prompt templates. The alignment with the threat behavior and technique templates is also completed based on the summary position of the tactical phase text.
\end{itemize}

%% file: 4_Evaluation.tex
\section{Evaluation}    
     In this section, we focus on evaluating the accuracy of AttacKG\textbf{+} as a CTI report parser for threat behavior graph extraction and technique identification, and its effectiveness as the CTI knowledge collector for technique-level intelligence aggregation. Meanwhile, the quality of the rewritten reports is assessed through manual analysis. Specifically, our evaluation is designed to answer the following research questions (RQs).
 
\begin{itemize}
    \item \textbf{RQ1:} %How accurate is CTEKG in extracting threat behavior graphs (entities and relations) from CTI reports?
    How does our system perform in terms of threat behavior graph extraction?
    \item \textbf{RQ2:} % \leftline{How accurate is CTEKG in identifying techniques }\\in CTI reports?
    What is the performance of the technique identification?
    \item \textbf{RQ3:} %How effective is CTEKG in aggregating technique-level intelligence across multiple CTI reports?
    How does the extracted AttacKG\textbf{+} look like? Does it reveal insightful information?
    \item \textbf{RQ4:} %\leftline{How effective is CTEKG rewrite in cyber threat} \\ intelligence processing?
    What is the quality of the rewriter module and does it significantly affect the downstream extraction?
\end{itemize}

\subsection{Experimental Setup}
     To evaluate AttacKG\textbf{+}, we collect 500 real event CTI reports from Cisco Talos Intelligence Group~\cite{Cisco}, Microsoft Security Intelligence Center~\cite{Microsoft}, and others.
     Meanwhile, we crawl the TTP Matrix template in MITRE for relevant definitions and examples and stored them in the Tactics-Techniques-Sub\_Techniques format hierarchy.
     We implement our prompt templates using the APIs of several commercial LLMs (such as GLM-4~\cite{zhipuai1}).
     
     To answer RQ1 and RQ2, we manually annotate entities, relations, and techniques as ground truth for the 15 reports that we collect. All the 15 reports describe real-world APT activities and contain relevant event elements.
     To answer RQ4, we recruit three graduate students who major in cyber security to manually assess the rewritten results of CTI reports, and we combine their assessment results to achieve a final evaluation.

\subsection{Evaluation Results}
%\textbf{\textbf{RQ1}}\textbf{\textbf{:}} How accurate is CTEKG in extracting threat behavior graphs from CTI reports?
\subsubsection{Effectiveness of threat behavior graph construction (RQ1)}

     Cyber threat events are composed of multiple threat atomic events in a temporal sequence.  These atomic events are represented in the threat behavior graph as interconnected triplets $(s, a, o)$, where $s$ and $o$ are harmonized as entity extraction, and $a$ and entity-entity relation are harmonized as relation extraction. Accurate construction of threat behavior graphs is crucial for identifying technique types and portraying threat events.  To evaluate the accuracy of AttacKG\textbf{+} construction, we use 15 CTI reports with good labeling mentioned earlier. We manually identify cyber threat entities based on the ontology model in the report and extract the relations between these entities based on domain knowledge and reporting logic.

     Given the real entities and relations in the report, we can compare AttacKG\textbf{+} with a state-of-the-art open-source CTI report parser in terms of precision, recall, and F-1 score of the Extractor~\footnote[3]{\url{https://github.com/ksatvat/EXTRACTOR}}.
     In particular, to ensure the fairness of the comparison, we take the following settings for EXTRACTOR and AttacKG\textbf{+}.
     For EXTRACTOR, we turn on the relevant optimization setting to ensure the extraction quality of entities and relations.

     Table~\ref{table-R1-R2} summarizes the statistical results of entity and relation extraction by AttacKG\textbf{+} and EXTRACTOR in 15 CTI reports.
     The experimental results are in the statistics: \textbf{Manual} identifies manually labeled real data. \textbf{Extractor} identifies the result data of EXTRACTOR. \textbf{AttacKG\textbf{+}} identifies the result data of AttacKG\textbf{+} run on the LLMs.
     Columns 2 through 7 show the counts of manually labeled ground truth and false\_negative(false\_positive) in the extraction of threat-related entities, and relations. Rows 18-20 represent the global \textbf{Precision}, \textbf{Recall}, and \textbf{F-1 Score}.
     
     It can be seen that in the results of entity extraction, although AttacKG\textbf{+} exhibits a slightly higher false\_positive rate in a small number of samples, the overall false\_positive rate performance is still lower than Extractor's extraction results.
     At the same time, AttacKG\textbf{+} shows a substantial lead in false\_negative rates, resulting in much higher overall accuracy and precision than Extractor.
     AttacKG\textbf{+} has both a low false\_negative and false\_positive rate in the relation extraction results.
     This shows that our method has made great strides in relation extraction and classification and that the overall F-1 score has improved significantly compared to current relation extraction schemes.

      The main reasons for the high rate of false negatives in EXTRACTOR are the aggregation of all non-Indicator of Compromise (IoC) entities of the same category during entity extraction, which leads to a loss of sequential information about the attack process, and the use of regular expressions for filtering during entity extraction, which is less sensitive to emerging threatening entities and cannot effectively recognize entity variants.  In contrast, AttacKG\textbf{+} can efficiently interpret emerging threat entities and their relationships, leveraging the broad-domain understanding capabilities of LLMs.

%\textbf{\textbf{RQ2:}} \leftline{How accurate is CTEKG in identifying techniques }\\in CTI reports? 
\subsubsection{Effectiveness of technique identification (RQ2)}

     To answer RQ2, we use AttacKG\textbf{+} to identify the techniques in 15 CTI reports and compared them with AttacKG.
     In the experiments of~\cite{AttacKG}, AttacKG~\footnote[4]{https://github.com/li-zhenyuan/Knowledge-enhanced-Attack-Graph} demonstrated excellent technique recognition compared to TTPDill. Therefore, we choose it as the comparison method.
     The main idea of AttacKG is to calculate the alignment score between the attack graph and the technique template using a graph alignment algorithm, which determines the type of technique included in the report based on a decision threshold. AttacKG uses default thresholds for the technique identification process in this paper.

\vspace{-1cm} 
\begin{table*}[h]
   \renewcommand\arraystretch{1.5}
    \centering
\caption{Accuracy of AttacKG\textbf{+} construction and technique identification.}
\scalebox{0.68}{
\label{table-R1-R2}         %这行要添加
\begin{threeparttable}  
\begin{tabular}{>{\raggedright\arraybackslash}p{3.3cm}|>{\raggedright\arraybackslash}p{1.38cm}>{\raggedright\arraybackslash}p{1.45cm}>{\raggedright\arraybackslash}p{1.6cm}|>{\raggedright\arraybackslash}p{1.38cm}>{\raggedright\arraybackslash}p{1.45cm}>{\raggedright\arraybackslash}p{1.6cm}|>{\raggedright\arraybackslash}p{1.38cm}>{\raggedright\arraybackslash}p{1.45cm}>{\raggedright\arraybackslash}p{1.6cm}}
    \toprule
 \multirow{2}{*}{CTI Reports}& \multicolumn{3}{c|}{Entities}& \multicolumn{3}{c|}{Relations}& \multicolumn{3}{c}{Techniques}\\ \cline{2-10}   
&  \;Manual&Extractor&AttacKG\textbf{+}& \;Manual&Extractor&AttacKG\textbf{+}&  \;Manual&AttacKG&AttacKG\textbf{+}\\ \midrule
 BRONZE&     \quad 13&\maRes{-13}{10} &\quad \maRes{-2}{9} &\quad \;8 &\maRes{-5}{18} &\; \maRes{-2}{9} &\quad \;4 &\maRes{-1}{18} &\; \maRes{-3}{4}\\ %\hline  
 Chat\_Mimi&     \quad 15&\maRes{-15}{9} &\quad \maRes{-5}{8} &\quad 10 &\maRes{-7}{15} &\; \maRes{-5}{4} &\quad\;4 &\maRes{-1}{7} &\; \maRes{-2}{1}\\ %\hline  
 North\_Korea&\quad 22 &\maRes{-19}{15} &\quad \maRes{-4}{5} &\quad \;9 &\maRes{-4}{22} &\; \maRes{-2}{4} &\quad\;7 &\maRes{-3}{23} &\; \maRes{-2}{2} \\ %\hline  
 Nitro\_Attacks&\quad 28 &\maRes{-28}{8} &\quad \maRes{-8}{5} &\quad 19 &\maRes{-6}{22} &\; \maRes{-7}{5} &\quad\;8 &\maRes{-5}{14} &\; \maRes{-3}{6} \\ %\hline  
 Moon\_Bounce&\quad 12 &\maRes{-12}{5} &\quad \maRes{-1}{10} &\quad 10 &\maRes{-6}{22} &\; \maRes{-5}{10} &\quad\;5 &\maRes{-2}{12} &\; \maRes{-3}{4} \\ %\hline  
 Stuxnet\_Under&\quad 24 &\maRes{-22}{21} &\quad \maRes{-8}{3} &\quad 18 &\maRes{-6}{31} &\; \maRes{7}{5} &\quad11 &\maRes{-8}{19}&\; \maRes{-5}{6} \\ %\hline  
 Stellar\_Particle&\quad 33 &\maRes{-32}{12} &\quad \maRes{-6}{5} &\quad 13 &\maRes{-5}{18} &\; \maRes{-5}{7} &\quad10 &\maRes{-10}{10} &\; \maRes{-1}{3} \\ %\hline  
 Prime\_Minister&\quad 19 &\maRes{-19}{10} &\quad \maRes{-5}{9} &\quad 12 &\maRes{-4}{12} &\; \maRes{-4}{3} &\quad12 &\maRes{-8}{11} &\; \maRes{-1}{1} \\ %\hline  
 Mustang\_Panda&\quad 37 &\maRes{-37}{10} &\quad \maRes{-9}{3} &\quad 19 &\maRes{-13}{28} &\; \maRes{-10}{7} &\quad12 &\maRes{-7}{22} &\; \maRes{-3}{9} \\ %\hline  
 Shuckworm\_APT&\quad 17 &\maRes{-16}{24} &\quad \maRes{-2}{11} &\quad \;9 &\maRes{-5}{18} &\; \maRes{-1}{8} &\quad\;7 &\maRes{-3}{9} &\; \maRes{-2}{4} \\ %\hline  
 C5\_APT\_SKHack&\quad 13 &\maRes{-11}{4} &\quad \maRes{-4}{4} &\quad \;9 &\maRes{-5}{18} &\; \maRes{-3}{1} &\quad\;5 &\maRes{-3}{17} &\; \maRes{-3}{4} \\ %\hline  
 Cisco\_Talos\_Bitter&\quad 17 &\maRes{-17}{10} &\quad \maRes{-9}{3} &\quad \;8 &\maRes{-5}{18} &\; \maRes{-3}{1} &\quad\;3 &\maRes{-2}{21} &\; \maRes{-1}{1} \\ %\hline  
 Log4Shell\_Rootkits&\quad 38 &\maRes{-36}{8} &\quad \maRes{-14}{7} &\quad 22 &\maRes{-13}{17} &\; \maRes{-10}{7} &\quad16 &\maRes{-12}{8} &\; \maRes{-9}{5} \\ %\hline  
 Cisco\_Talos\_Iranian&\quad 14 &\maRes{-14}{8} &\quad \maRes{-3}{7} &\quad \;6 &\maRes{-3}{19} &\; \maRes{-3}{2} &\quad\;4 &\maRes{-2}{9} &\; \maRes{-3}{1} \\ %\hline  
 Asylum\_Ambuscade&\quad 21 &\maRes{-21}{10} &\quad \maRes{-9}{3} & \quad 11 &\maRes{-6}{24} &\; \maRes{-4}{3} &\quad\;4 &\maRes{-1}{16} &\; \maRes{-1}{3} \\ \midrule  
 \textbf{Overall  Precision}&\;\textbf{1.000} &\quad \textbf{0.046} &\quad \textbf{0.668} &\;\textbf{1.000} &\; \textbf{0.221} &\quad\textbf{0.601} & \;\textbf{1.000} &\; \textbf{0.179} &\quad\textbf{0.545}\\ %\hline  
 \textbf{Overall Recall}&\;\textbf{1.000} &\quad \textbf{0.034} &\quad \textbf{0.732} &\;\textbf{1.000} &\; \textbf{0.472} &\quad\textbf{0.647} &\;\textbf{1.000} &\; \textbf{0.458} &\quad\textbf{0.588} \\ %\hline  
 \textbf{Overall  F-1 Score}&\;\textbf{1.000} &\quad \textbf{0.039} &\quad \textbf{0.698} &\;\textbf{1.000} &\; \textbf{0.301} &\quad\textbf{0.623} &\;\textbf{1.000} &\; \textbf{0.258} &\quad\textbf{0.566}\\ \bottomrule 
\end{tabular}

\begin{tablenotes}\footnotesize
    \item[1] Accuracy of threat behavior graph construction and technique identification in 15 CTI reports.
    \item[2] Columns 2--10 present the  ground-truth and \textbf{false negative/positive} in extracting entities, relations, and techniques.
    \item[3] Rows 18--20 present the overall Precision, Recall, and F-1 Score.
\end{tablenotes}
\end{threeparttable}
}
\end{table*}
\vspace{-0.6cm} 

     We evaluate the effectiveness of AttacKG\textbf{+}'s and AttacKG's technique identification across 15 reports, which have been manually labeled with the ground-truth technique. 
    The results of the technique identification are summarized in Table~\ref{table-R1-R2}.
    Columns 8 to 10 display the counts of manually labeled ground truth and false\_negative(false\_positive) in the extraction of technique identification.

It is evident that both AttacKG\textbf{+} and AttacKG maintain low false-negative rates, with AttacKG\textbf{+} exhibiting greater stability. However, AttacKG tends to incur a significant number of false positives, averaging 14.4 per report. 
AttacKG\textbf{+} surpasses AttacKG with a precision rate that is 0.366 higher and a recall rate that is 0.130 higher. Consequently, AttacKG\textbf{+}'s F-1 score is roughly double that of AttacKG. 
The experimental data demonstrates that AttacKG\textbf{+} outperforms previous methods in technique identification tasks.

     Specifically, since AttacKG does not have a text filtering 
 rewriter, we find that AttacKG is less effective in extracting raw report text when dealing with complex structures.
     This is because the report text contains a significant amount of background information unrelated to the attack process. Consequently, AttacKG generates numerous discrete subgraphs that fail to represent the full sequence of the event.

\subsubsection{AttacKG\textbf{+} statistics and analysis (RQ3)}

     To answer RQ3, we use AttacKG\textbf{+} to extract 500 pieces of cyber threat intelligence originating from different sources on tactics, technologies, and threat-related entities and relations.
     Table~\ref{table3}  shows a list of the top five distributions of the above extracts in the report.

     It can be found that the Initial Access tactic is the most common among all tactics, consistent with the notion that malicious access is prioritized in threat events. The Top 5 tactics distribution also indicates that CTI reports tend to focus more on the Command and Control stage. The top 5 techniques are similar to the top TTP matrices manually generated by PICUS~\cite{picus}. All 14 tactics are covered in all CTI reports, with an average of 14.61 TTP techniques and 40.7 threat entities per report. By analyzing the tactical distribution of the rewritten CTI reports and conducting manual validation, we identified a new classified pattern for the CTI reports.

 \vspace{-0.9cm} 
\begin{table*}[htbp]
    %\caption{Effectiveness of Threat Intelligence Extraction from 500 CTI Reports.}
    \caption{Effectiveness of threat intelligence extraction from 500 CTI reports.}
    \label{table3}
       \centering
\scalebox{0.7}{

    \begin{threeparttable}    
 
    \renewcommand\arraystretch{1.5}
\begin{tabular}
   {>{\centering\arraybackslash}m{3.4cm}|>{\centering\arraybackslash}m{1.2cm}|>{\centering\arraybackslash}m{1.6cm}|>{\centering\arraybackslash}m{1.2cm}|>{\centering\arraybackslash}m{2cm}|>{\centering\arraybackslash}m{1.2cm}|>{\centering\arraybackslash}m{3cm}|>{\centering\arraybackslash}m{1.2cm}}
   \toprule
   Tactics&  Occur&  Techniques&  Occur&  Entities&  Occur&  Relations& Occur\\ 
   \midrule
   
         Initial Access
&1467  &  T1566
&663  &  Attacker
&768  &  Use
&2062 \\ %\hline 
         Command and Control
& 1221 &  T1071
&388  &  Malware
&399  &  Target
&462 \\ %\hline 
         Execution
& 1217 &  T1059
&305  &  C \& C Server
&369  &  Communicate With
&240 \\ %\hline 
         Defense Evasion
&1017  &  T1547
&261  &  Campaign
&107  & Execute
&217 \\ %\hline 
         Persistence
&731  &  T1190
&200  &  Attack
&94  &  Host
&191 \\ \midrule

         \textbf{All}&\textbf{8411}  &  \textbf{All}&\textbf{7305}  &  \textbf{All}&\textbf{20350}  &  \textbf{All}&\textbf{10175} \\ \bottomrule
    \end{tabular}
\begin{tablenotes}
\footnotesize 
\item [1] This table shows the top 5 tactics, techniques, entities, and relationships from 500 CTI reports.
\end{tablenotes} 
\end{threeparttable} 
}
\end{table*}
\vspace{-0.9cm}

\begin{itemize}[leftmargin=*]
      \item \textbf{High-Tactic}. Suppose the report has numerous tactical phases, it is likely to be a historical record of APT operations or a profile of the APT organization.  This is because APT operations or organizations often require a combination of tactics/techniques to achieve the purpose of the intrusion.
     \item \textbf{Low-Tactic}. Suppose the report contains fewer types of tactics, it is likely a malware introduction or vulnerability analysis report. This is because malware or exploits commonly serve as a single component of a larger intrusion, appearing as a part of a specific technique, which is why the report would focus on just one tactic. 
     \item   \textbf{Non-Tactic}. Suppose the main body of the report doesn't cover the tactical phase, it usually turns into a social news piece about a threat event or a step-by-step threat overview. This is because these reports usually don't focus on the tactical and technical aspects of the incident, but instead provide a general summary of the threat situation.
\end{itemize}

    Realizing that this is a new perspective for categorizing the types of threat reports, we categorize  similar types by reported keywords in constructing the datasets Re-CTI and CTI-TE, which is more helpful in filtering out the tactical cyber threat intelligence with temporal characteristics describing the process of the intrusion and serves as a new categorization criterion for all types of CTI.

%\textbf{\textbf{RQ4}}\textbf{:} How effective is the CTEKG rewrite in cyber threat intelligence processing?
\subsubsection{Quality and impact of the rewriter (RQ4)}

     The implementation of tactical report rewriting is dependent on the ability to understand a combination of multiple texts, including tactical definitions and threat reports, through the capabilities of LLMs. 
     Since there is no related work on the quality of threat intelligence rewriting for LLMs.
     We design a manual evaluation mechanism to assess the quality of rewritten tactical reports from 15 rewritten reports in RQ1 and RQ2 in terms of two levels: delineation accuracy, and information consistency.

     We use "sentence" as the unit of assessment, and each category is categorized into four levels (Level 1-4) with the following assessment criteria.

 \vspace{-0.6cm} 
\begin{figure}[htbp]
    \centering
    \includegraphics[width=0.8\linewidth]{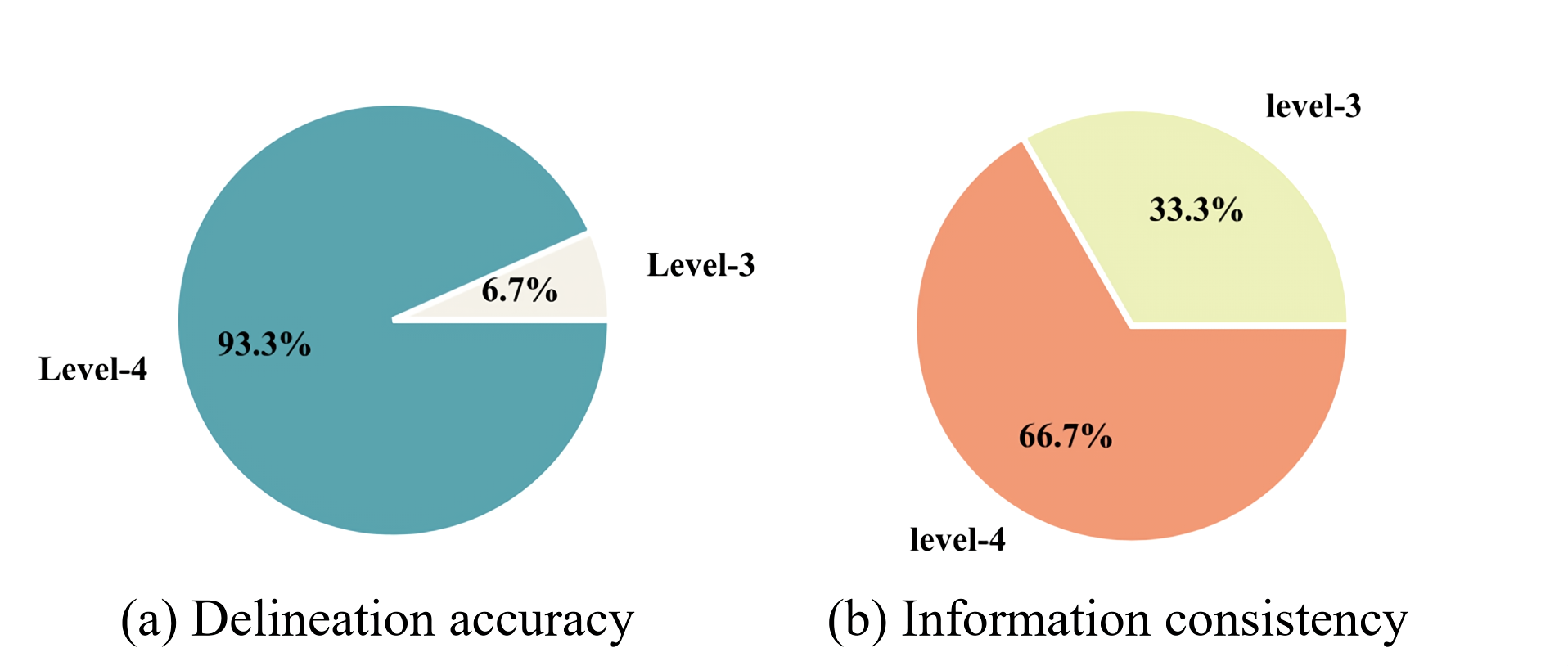}
    \caption{Manual assessment results of the report (tactical rewrite version).}
    \label{fig5}
\end{figure}
\vspace{-0.6cm}

\begin{itemize}[leftmargin=*]
      \item \textbf{Delineation Accuracy}. Evaluate whether the content of the rewritten text is in the correct tactical phase.
\textbf{Level-1}.The results of the tactical stage division differ significantly from the original text, with low accuracy. $(\leqslant 30 \%)$
\textbf{Level-2}.The results of the tactical stage division are partially in line with the original text, with moderate accuracy.$(30 \%<x \leqslant 60 \%)$
\textbf{Level-3}.The results of the tactical stage division are largely consistent with the original text, with good accuracy. $(60 \%<x \leqslant 90 \%)$
\textbf{Level-4}.The results of the tactical stage division are fully consistent with the original text, with very good accuracy. $(>90 \%)$
      \item \textbf{Information Consistency}. Evaluate whether the content of the rewritten text matches the semantics of the original report.
\textbf{Level-1}.The threat intelligence report's rewritten meaning is very different from the original text, with low consistency. $(\leqslant 30 \%)$
\textbf{Level-2}.The threat intelligence report's rewritten meaning partially matches the original text, with medium consistency.$(30 \%<x \leqslant 60 \%)$
\textbf{Level-3}.The threat intelligence report's rewritten meaning is largely consistent with the original text, with high consistency. $(60 \%<x \leqslant 90 \%)$
\textbf{Level-4}.The threat intelligence report's rewritten meaning is fully consistent with the original text, with very good consistency. $(>90 \%)$
\end{itemize}

     Figure~\ref{fig5} summarizes the results of the quality assessment. It is evident that the rewritten reports by the LLMs exhibit an impressive level of categorization accuracy, with 14 out of 15 reports demonstrating a delineation accuracy of Level-4. This indicates that the LLMs possess a strong understanding of content categorization. However, there is a slight inconsistency in the information presented in the report compared to the original text, due to the limitations of the model's quality and the corpus. Overall, the LLMs have shown improvement in rewriting threat reports, but further enhancements may be necessary to overcome the technical restrictions of the model.

     The reasons for choosing manual evaluation over ablation in this paper are as follows: 1) Generative modeling mechanism. The rewritten report, generated by a generative language model, contains categorized and rewritten entities and relations that cannot be compared with the original text in the same benchmark. 2) Threat event extraction requirements.   Full-text-based content summary rewriting is necessary, as it involves both a full-text security background and a focus on the full-text attack implementation process, with a large amount of context information loss when only selecting part of the original report.

\subsection{Case Study}
\label{sec:case study}
     This subsection discusses how AttacKG\textbf{+} can be employed in real-world security tasks through a case study. Security practitioners and researchers often perform in-depth analysis for occurred threat events, the process of which usually includes two perspectives:
     First, they need to extract the subject that is closely related to the attack process from the complex event information and sift out redundant information. 
     Second, they demand to bridge the knowledge gap between CTI reports and real attacks to identify the techniques that are potentially used.
     % We show its utility in the scenario of threat event reconstruction.

\begin{figure}[htbp]
    \centering
    \includegraphics[width=0.95\linewidth]{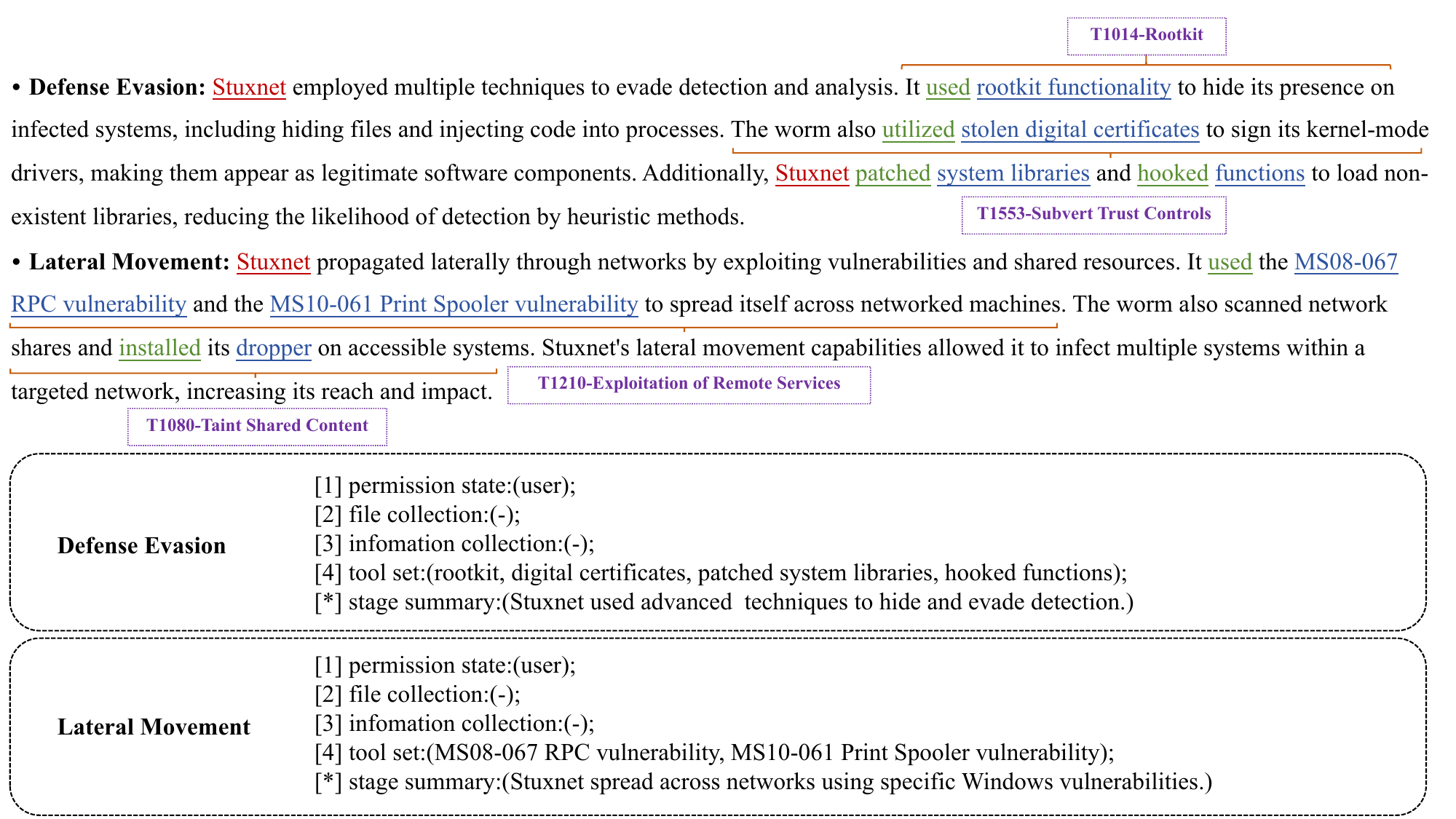}
    \caption{Example of complex threat event extraction.
    % For CTI reports, three levels of information are of interest: 
    1) the distribution of threat entities and relations. 2) the use of cyber attack techniques. 3) the stage state summary situation, containing permissions, files, tools, and information.
}
    \label{Motivation}
\end{figure}

     Specifically, AttacKG\textbf{+} employs structured knowledge to parse events, thus helping to understand and reconstruct the threat events involved to achieve the above two objectives. As shown in Figure.~\ref{Motivation}, Based on the idea of AttacKG\textbf{+}, for a structured description of an attack process or incident, three levels of threat event information are required to be extracted from rewritten CTI reports.     
     We develop a visualization interface for AttacKG\textbf{+} using pyvis~\footnote[5]{\url{https://github.com/WestHealth/pyvis}}. The visualization for example AttacKG\textbf{+} is presented in Appendix~B.

     Such as Figure.~\ref{fig:sub1}, taking the attack against SK Communications (C5 APT SHack) as an example, according to the extracted  AttacKG\textbf{+} provides structured knowledge about threat event scenarios.
     % enabling security analysts to have richer process information when reviewing threat events and making it easier to mine potential features and commonly share characteristics across events.
     From Figure.~\ref{fig:sub1}, we can easily identify six techniques and five tactics involved in the attack, including \textit{T1195-Supply Chain Compromise} for tactic Initial Access, \textit{T1059-Command and Scripting Interpreter} for tactic Execution, \textit{T1070-Exploitation for Execution} for tactic Defense Evasion, and so on.
     We can then infer the system environment in which the intrusion activity occurred based on the technologies and entities involved in the threat event.
     Specifically, the installation of malware named nation.exe implies that the victim's system environment is Windows. Execution using the malware 'Backdoor.Agent.Hza' indicates the presence of a version of the Trojan in effect or a setup flaw in the system.
     After confirming the environment in which the threatening event occurred, by combining the state summary, it can be found that the attacker's main objective is to steal a massive amount of user information from SK Communications.
     All in all, the AttacKG\textbf{+} provided the necessary information that can be used as a basis for restoring the threat event.

\vspace{-0.8cm} 

\vspace{-0.8cm}

%% file: 5_Related_Work.tex
\section{Related Work}
In this section, we briefly review the related works from two branches: 1) CTI Report Extraction, 2) LLMs for Cyber Security.

\textbf{CTI Report Extraction}
Current CTI extraction work consists of three aspects in terms of information granularity: CTI report extraction, unstructured information extraction, and TTP Identification. Existing OSCTI collection and management systems~\cite{Threatcrowd,Abuse,PhishTank,OpenPhish,OpenCTI,AlienVault,IBM} mainly focus on IoCs sharing~\cite{Acing}, such as the name and type of a malicious file, the number and name of a malicious process, and the hash value of a malware sample.
When knowledge is embedded in a large amount of text, it is difficult to be used effectively. To solve this problem, relevant methods for extracting threat information from unstructured text have emerged~\cite{SemiAutomated,Ghazi,Ghaith,CyberEntRel}.
Further, in addition to extracting cyber threat entities and relationships from unstructured data, identifying the technology patterns that are used in the threat scenarios enables a more comprehensive perception of the intrusion process, and the TTP matrix proposed by MITRE has become the mainstream reference standard for technology pattern labeling~\cite{TTPDrill,Ayoade,SeqMask}.

\textbf{LLMs for Cyber Security}
Currently, scholars have been extensively exploring the use of large language models in the field of cybersecurity.
Maxime et al.~\cite{wrsch2023llms} use the LLMs to extract relevant knowledge entities from cybersecurity-related texts.
Mohamed et al.~\cite{ferrag2023revolutionizing} found that FalconLLM shows great potential in performing the identification of complex patterns and complex vulnerabilities.
Maria et al.~\cite{rigaki2023cage} use pre-trained LLMs as agents in a cybersecurity network environment, focusing on their utility in the sequential decision-making process.
Andrei et al.~\cite{kucharavy2023fundamentals} provide a systematic overview of the history, current status, and impact of generative language modeling, including its principles, capabilities, limitations, and prospects.
Benjamin~\cite{kereopayorke2023building} qualitatively analyzes SME case studies and quantitatively assesses LLM metrics in cybersecurity applications.

%% file: 6_Conclution_and_Future_work.tex
\section{Conclusion and Future Work}
%\textbf{Conclusion}.
% In this work, we emphasized three crucial characteristics of threat events and proposed a Complex Threat Event Knowledge Graph (CTEKG) containing the behavior graph, TTP labels, and state summary.
In this work, we proposed a fully automatic LLM-based framework (AttacKG\textbf{+}) to construct attack knowledge graphs, 
based on the LLMs' exceptional capabilities in both language understanding and zero-shot task fulfillment.
% based on the powerful contextual understanding capability and the convenient instruction tuning capability of LLMs.
Meanwhile, given the superior capabilities of LLMs, we have upgraded the knowledge scheme used by current research, by introducing three levels of threat knowledge: threat behavior, TTP labels, and state summary.
As a by-product, we constructed two threat intelligence datasets, Re-CTI and CTI-TE.
Finally, we conducted extensive experiments to evaluate the effectiveness of our proposed method, and the results demonstrate the great potential of LLMs for attack knowledge graph construction.

%\noindent\textbf{Future Work}.
     Although the motivation and design of AttacKG\textbf{+}  are valuable, the construction of the CTI extraction pipeline still has some limitations.
     1) First, the current construction only consider textual input while ignoring multimodal information, such as figures and charts. 
     The visual elements in CTI reports, such as intrusion processes and software architecture, contain valuable information. However, exist research primarily focused on natural language, limiting the exploration of these other modalities. Therefore the use of multimodal threat information in future work will be able to enhance LLM's ability to analyse CTI reports.
     2) Second, the currentl LLMs still do not clearly understand user requirements in some scenarios, which affects the quality of the output of the task results. Therefore, it is important to better align the LLMs with the values of the expert's approach to the problem. When the LLMs' understanding of requirements improves, we can be able to perform more complex tasks with better results.

%% file: 7_Appendix.tex
\section{Appendix}
\label{Appendix}
\appendix
%% \label{}
% \onecolumn
% \section{Appendix. A}

\section{Templates of AttacKG\textbf{+} Extraction}
In the AttacKG\textbf{+} extraction process, four specific prompts are used: the rewriting prompt, triplet extractor prompt, technique identifier prompt, and stage summarizer prompt. Detailed descriptions of each prompt's design are available at the provided link\footnote[6]{\href{https://anonymous.4open.science/r/CTKEG_Appendix-19DC/}{\url{https://anonymous.4open.science/r/CTKEG_Appendix-19DC/}}}. 

% The template of the rewriting prompt is shown in Table~\ref{tab:template_prompt_rewriting}.
\textbf{Rewriting Prompt. } The rewriting prompt aims to standardize threat intelligence reporting by restructuring the original content according to the 14 tactical categories of the TTP matrix, ensuring the preservation of temporal sequences, entity classifications, and relational categories.
All tactics, techniques, entities, and relations are included in the prompt to instruct the LLMs to preserve the critical information as much as possible.

% The template of the triplet extractor prompt is shown in Table~\ref{tab:template_prompt_quadruples_extractor}.
\textbf{Triplet Extractor Prompt. } The triplet extractor prompt is used to extract "subject-action-object" triplets and entity-entity relations in the rewritten tactic paragraph. To reduce the expenses associated with the commercial LLMs, multiple tactic paragraphs with the same tactic are integrated into a single prompt.
Overly lengthy prompts can compromise the accuracy of information extraction and may overlook critical details. Therefore, it is recommended to limit the number of paragraphs processed concurrently to a maximum of 10 to ensure optimal results.
% Intuitively, these tactic labels of the triplets are the same ones that the rewritten paragraphs share.

\balance

% The template of the technique identifier prompt is shown in Table~\ref{tab:template_prompt_technique_labeler}.
\textbf{Technique Identifier Prompt.} Technique identifier is used to add technique labels for extracted triplets. 
triplets that are assigned an identical tactic label are also associated with a uniform candidate technique label according to MITRE. 
This association is delineated in two rules: Rule 1 enumerates the possible techniques associated with a given tactic, while Rule 2 explains each technique's specific meaning.
The format presents the extracted triplet followed by the corresponding context paragraph. It is structured as "serial number: subject; relation; object; context paragraph".

% The template of the state summarizer prompt is shown in Table~\ref{tab:template_prompt_stage_summarizer}.
\textbf{State Summarizer Prompt. } State summarizer prompt is used to summarize the resources the attacker gets when the actions in one tactic stage are finished.
The resources include permission state, file collection, information collection, and tool set. 
The permission state refers to the level of system access that attackers have managed to acquire.
% , typically starting with user-level privileges by default. 
File collection encompasses the variety of file types, such as .txt, .doc, and .jpg, that attackers exfiltrate from the compromised systems.%which often contain sensitive data.% 
 Information collection goes a step further by detailing the extraction of highly confidential data by attackers, including but not limited to passwords, personal identities, and account details. 
 % that can have dire consequences if misused.
The tool set contains the tools in the paragraph used by the attacker to achieve their purpose.
% Additionally, a concise overview of the tactic stage is presented to clarify the attacker's actions.

\section{Cases of AttacKG\textbf{+} Visualization}
Appendix B presents a series of examples that demonstrate the visualization of the extracted attack knowledge graphs, which are derived from cyber attack incidents described in CTI reports. 
The visualization is divided into two main sections: the threat behavior graph (nodes are shown in green and blue) and the tactic and technique labels of the event nodes (highlighted in red and yellow), And the state summary we store in the form of an index in the data file available for querying.
In the AttacKG\textbf{+}, triangular green nodes represent event nodes which are actions carried out by entities. Blue nodes appear in two shades: lighter blue nodes indicate the subjects that perform actions, while darker blue nodes denote the objects to which the actions are directed.
The cyber events illustrated include C5 APT SKHack, Chat Mimi, MoonBounce, and Asylum Ambuscade. An in-depth discussion of the C5 APT SKHack event is provided in Section \ref{sec:case study}.

% \end{figure}
\onecolumn

\begin{figure}[htb] % 'h' for here, 't' for top, 'b' for bottom

  \centering % 居中显示

  % 子图1
  \begin{subfigure}[b]{\textwidth} % 子图宽度为文本宽度的30%
    \includegraphics[width=\textwidth]{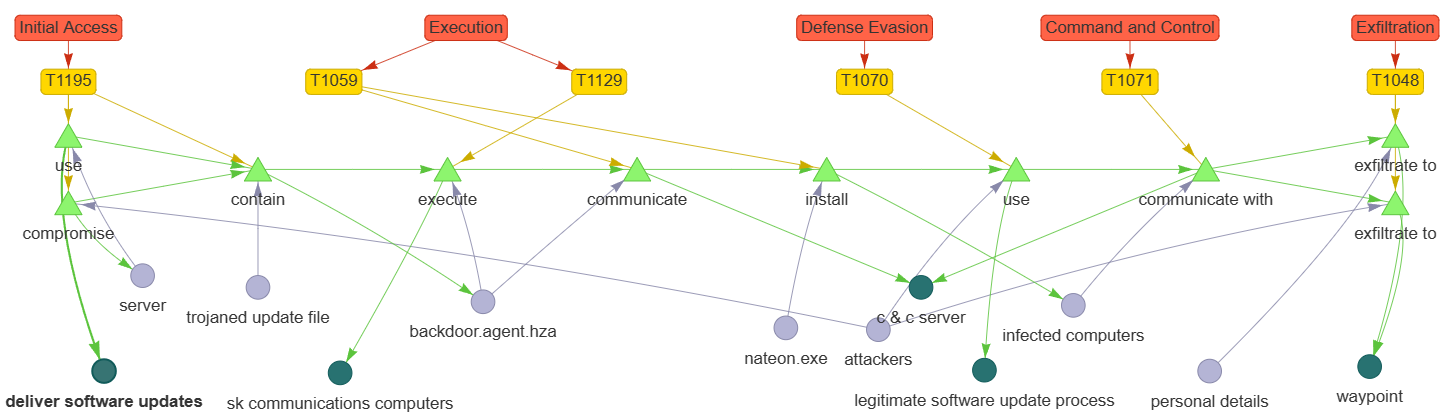}
    \caption{C5 APT SKHack}
   
    \label{fig:sub1} % 子图引用标签
  \end{subfigure}
  \hfill % 水平填充，确保子图之间的间距
  \\
  % 子图2
  \begin{subfigure}[b]{\textwidth}
    \includegraphics[width=\textwidth]{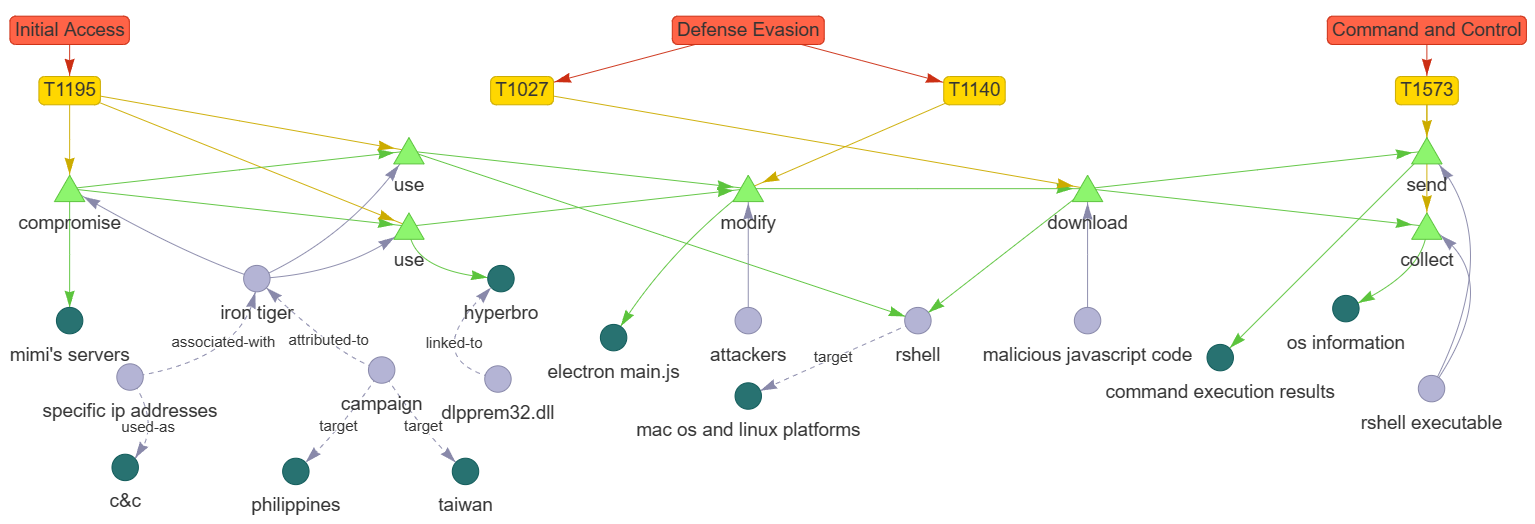}
    \caption{Chat Mimi}
    \label{fig:sub2}
  \end{subfigure}
  \hfill
  \\
  % 子图3
  \begin{subfigure}[b]{\textwidth}
    \includegraphics[width=\textwidth]{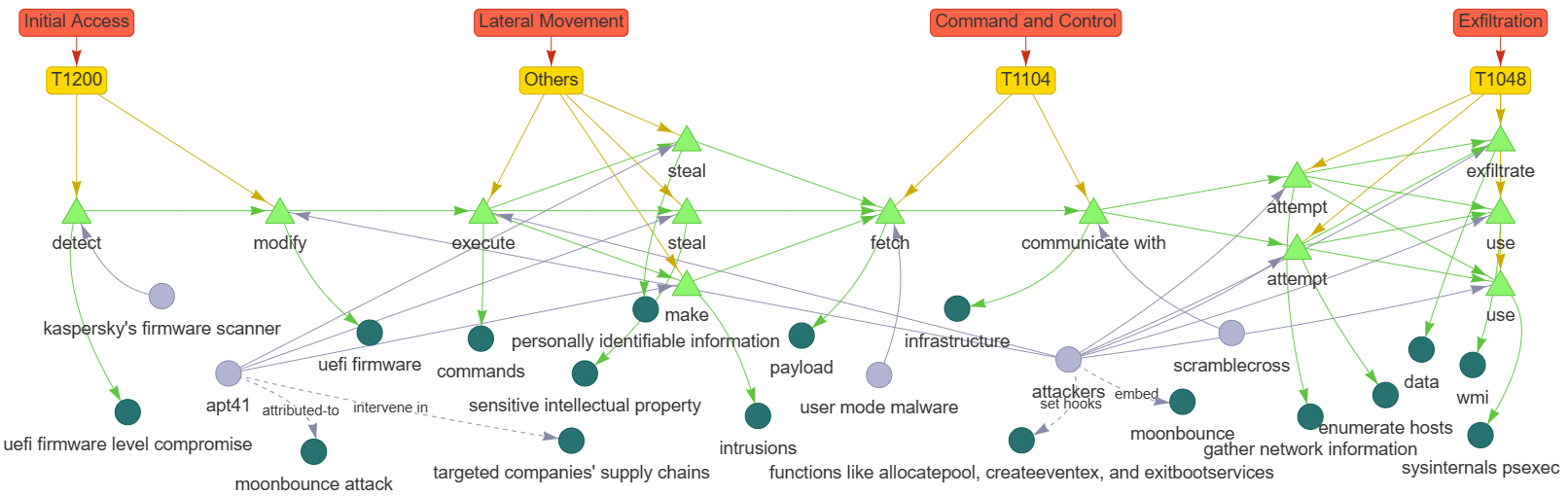}
    \caption{Moon Bounce}
    \label{fig:sub3}
  \end{subfigure}
    \\
  % 子图4
  \begin{subfigure}[b]{\textwidth}
    \includegraphics[width=\textwidth]{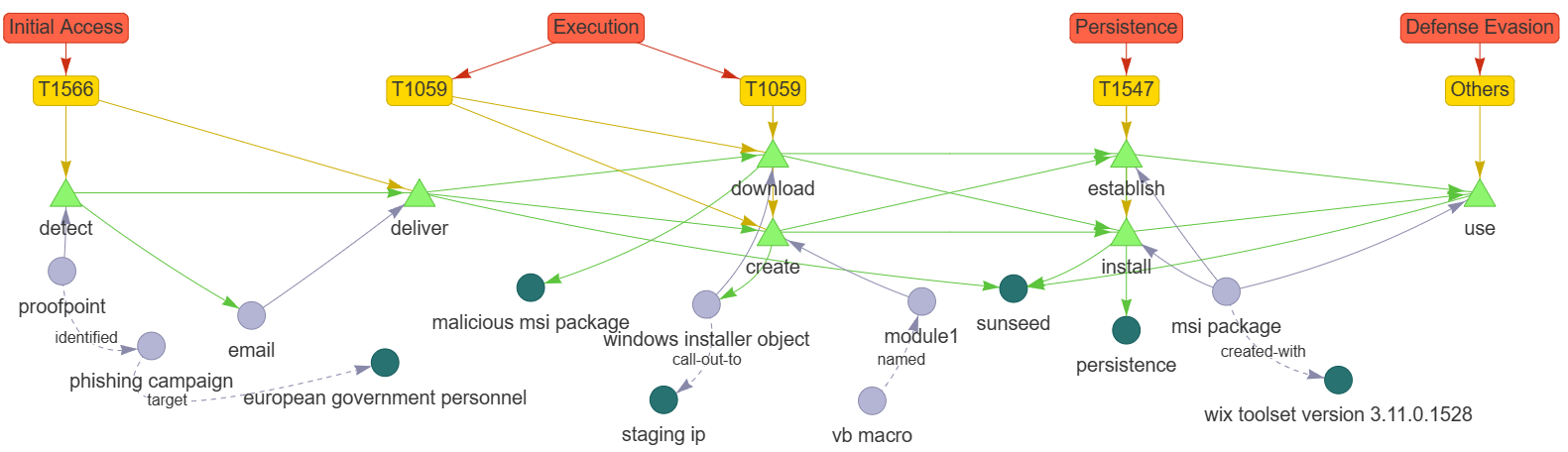} % 替换为实际的图片文件
    \caption{Asylum Ambuscade} % 子图标题
    \label{fig:sub4}
  \end{subfigure}
  \caption{Examples of the visualization of AttacKG\textbf{+}} % 总图的标题
  \label{fig:visualizations_3} % 总图的引用标签

\end{figure}